\documentclass{article}
\usepackage[utf8]{inputenc}
\usepackage{fullpage}
\usepackage{appendix}
\usepackage{graphicx}
\usepackage{amsmath}
\usepackage{amssymb}
\usepackage{amsthm, amsfonts}
\usepackage{algorithm, algorithmic}
\usepackage{subfig}
\usepackage{hyperref}
\usepackage{float}
\usepackage{tikz}
\usepackage[numbib]{tocbibind}
\usepackage[numbib]{tocbibind}
\usepackage{setspace}
\usepackage{etoolbox}
\apptocmd{\thebibliography}{\raggedright}{}{}

\newtheorem{proposition}{Proposition}

\newtheorem{corollary}{Corollary}[proposition]

\title{Liquidity Provider Returns in Geometric Mean Markets}
\author{Alex Evans\footnote{alex@placeholder.vc}}
\date{June 2020}

\begin{document}
\maketitle

\begin{abstract}
Geometric mean market makers (G3Ms), such as Uniswap and Balancer, comprise a popular class of automated market makers (AMMs) defined by the following rule: the reserves of the AMM before and after each trade must have the same (weighted) geometric mean. This paper extends several results known for constant-weight G3Ms to the general case of G3Ms with time-varying and potentially stochastic weights. These results include the returns and no-arbitrage prices of liquidity pool (LP) shares that investors receive for supplying liquidity to G3Ms. Using these expressions, we show how to create G3Ms whose LP shares replicate the payoffs of financial derivatives. The resulting hedges are model-independent and exact for derivative contracts whose payoff functions satisfy an elasticity constraint. These strategies allow LP shares to replicate various trading strategies and financial contracts, including standard options. G3Ms are thus shown to be capable of recreating a variety of active trading strategies through passive positions in LP shares.

\end{abstract}

\section{Introduction}

Decentralized Finance (DeFi) consists of a set of protocols and applications that provide automated financial services through smart contracts. At the time of writing, it is estimated that nearly 1 billion USD~\cite{defipulse} is being utilized by DeFi systems. DeFi applications often employ automated market makers (AMMs) to offer standard financial services such as trading~\cite{curve_2019} and lending~\cite{compound2019}, as well as less conventional products such as perpetual swaps~\cite{futureswap_2020} and flash loans~\cite{flash_loans}. 

Among AMM designs, geometric mean market makers (G3Ms) are most common to Decentralized Exchanges (DEXs) such as Uniswap~\cite{uniswap} and Balancer~\cite{balancer_wp}. In G3Ms, liquidity providers deposit assets into the reserves of a smart contract. This contract permits third parties to submit trades against supplied reserves, executing a trade only if the weighted geometric mean of reserves after the trade is equal to the one before. In exchange for supplying reserves to the contract, liquidity providers are issued liquidity pool (LP) shares in proportion to their contributions. LP shares may be redeemed for a proportional share of the pool's reserves at any time. The marginal prices offered by G3Ms are known to closely track prices on more liquid trading venues~\cite{AC20}. This occurs because arbitrageurs are incentivized to respond to price fluctuations by submitting trades that rebalance reserves to target weights~\cite{balancer_wp}. This activity is akin to automated Exchange Traded Fund (ETF) rebalancing.

\paragraph{Numerical example.} While a formal definition of G3Ms is provided in \S\ref{sec:g3ms}, it is instructive to examine a simple numerical example first. Consider two investors who each add 5 units of asset $A$ and 5 units of asset $B$ to a G3M that assigns weights $w_A=1/3$ to asset $A$ and $w_B=2/3$ to asset $B$. The weighted geometric mean of reserves is then $10^{1/3}10^{2/3}=10$. If a trader sends 1 unit of asset $A$ to the smart contract and demands 5 units of asset $B$ in exchange, the trade will be rejected, as the post-trade weighted mean would be $11^{1/3}5^{2/3} \neq 10$. However, a trade that adds 1 unit of asset $A$ in exchange for 0.466 units of asset $B$ will be accepted, as $11^{1/3}9.534^{2/3} = 10$. Clearly, the price offered by the G3M in this trade is 1 unit of asset $A$ which is added to the LP, for 0.466 units of asset $B$ which is removed from the LP. This price depends only on the pre-trade reserves $R_A=R_B=10$ and the weights $w_A=1/3$, $w_B=2/3$. After the trade, each investor's LP shares are redeemable for half of the reserves, namely 5.5 units of asset $A$ and 4.767 units of asset $B$. We refer to the total value of reserves that the LP shares can be redeemed for as their ``payoff." 

The marginal price offered by the G3M is the amount of asset $B$ a trader receives in exchange for a small quantity of asset $A$ (and vice versa). When the marginal price offered by the G3M doesn't reflect the true market price, an arbitrage opportunity results to adjust the reserves of the G3M. For example, consider again the case where the LP consists of 10 units of asset $A$ and 10 units of asset $B$. If the price of asset $B$ is $S_B=\$ 2$ and the price of asset $A$ is $S_A= \$1$, then the LP holds $\$30$ worth of assets, of which $1/3$ is held in asset $A$ and $2/3$ in asset $B$. This allocation agrees with the respective weight of each asset, $w_A=1/3$ and $w_B=2/3$. If the external price of asset $B$ drops to $S_B'=\$ 1$, then, to restore the allocation so that $1/3$ of the LP's value is in asset $A$ and $2/3$ in asset $B$, a trader sends $2.6$ units of asset $B$ to the smart contract. In exchange, the contract sends $3.7$ units of asset $A$ to the trader, maintaining the geometric mean of $(10-3.7)^{1/3}(10+2.6)^{2/3} = 10$. The trader thus makes an arbitrage profit of $3.7-2.6=1.1$. After the trade, the reserves are updated to $R_A=6.3$ in asset $A$ and $R_B=12.6$ in asset $B$. The total value held in the LP is $S_AR_A+S_B'R_B=\$6.3+\$12.6=\$ 18.9$, of which $6.3/18.9=1/3$ is held in asset $A$ and $2/3$ in asset $B$ (again corresponding to the respective weights of the two assets). One can check that sending any amount of asset $B$ to the G3M other than $2.6$ results in lower arbitrage profits for the trader. For example, sending 2 units would yield a profit of 1.06, while sending 3 units would yield a profit of 1.08. This insight is formalized in~\cite{AC20,balancer_wp}, which show that adjusting the reserves 
so that $1/3$ of the LP's value is held in asset $A$ and $2/3$ in asset B maximizes arbitrage profits for the trader. Traders are therefore incentivized to respond to price changes by rebalancing the reserves of the G3M to match the target weights.

\paragraph{G3Ms in practice.} The most well-studied examples of G3Ms are the Uniswap and Balancer protocols. Uniswap exclusively supports LPs consisting of two assets whose reserves are equally weighted. This simplifies the geometric mean to a ``constant product rule" that allows traders to perform any trade that preserves the product of reserves. The simplicitly and apparent effectiveness of Uniswap has spurred other applications to adopt the constant product rule~\cite{clabscelo, futureswap_2020, mcdex, dimaggio_2019}. 

Balancer generalizes the constant product formula by allowing pools of multiple assets as well as configurable weights. Balancer also supports dynamic weights that can be updated according to a set of rules~\cite{smartpools}. For example, this allows the LP to gradually decrease its exposure to an asset over time~\cite{lbps} or to adjust weights to favor assets that exhibit lower volatility~\cite{piedao}. 

As of this writing, Uniswap has nearly 60 million USD in reserves and facilitates 10 million USD in daily trading volume, while Balancer has approximately 30 million USD in reserves and facilitates nearly 1 million USD in daily trading volume~\cite{ defipulse,uniswapinfo,dune_dexs,dune_balancer}. Amid growing interest in G3Ms, DeFi lending platforms have started accepting LP shares as collateral for secured loans~\cite{aave_lps}. As G3Ms are attracting larger amounts of capital and their LP shares are being used in increasingly complex financial transactions, there is a rising need for a unified framework to study the return and price characteristics of LP shares in G3Ms.

\paragraph{Prior work.} AMMs have been widely studied since the the introduction of the popular logarithmic market scoring rule~\cite{hanson_lmsr}. The present paper focuses on LP share returns in G3Ms, which are a popular class of AMMs pioneered by~\cite{uniswap,balancer_wp}. The most relevant prior work in this context is that of~\cite{ACC19,AC20,clark}. Specifically,~\cite{ACC19} derives returns and prices of  LP shares in Uniswap, which consists of two equally-weighted assets, while~\cite{AC20} derives an expression for LP share returns in constant-weight G3Ms consisting of more than two assets. For Uniswap,~\cite{clark} replicates LP share payoffs with the spanning formula of \cite{Carr98towardsa} and demonstrates approximate hedging techniques using portfolios consisting of Uniswap LP shares and positions in futures contracts.

\paragraph{Overview.} This paper studies LP share returns in generalized no-fee G3Ms. The static-weight payoff results in~\cite{ACC19} and~\cite{AC20} are extended to G3Ms with dynamic weights. In a parametric setting, the no-arbitrage prices of LP shares are shown to follow directly from these payoff solutions. The resulting prices can be used to analyze certain properties of LP share returns, such as per-trade losses and value leakage from volatility. This paper also shows how to use LP shares to replicate target payoffs. We show that setting the weight of a G3M equal to the elasticity of a given payoff function ensures that the LP shares replicate the payoff. The elasticity of a derivative's payoff is defined as the percent change in the derivative's value per percent change in the price of the underlying asset it references. For differentiable payoff functions that have elasticity between zero and one, the resulting hedges are exact and do not depend on the model one uses for the underlying asset price. Replication is also studied under more general assumptions by utilizing parametric hedges. G3M LPs are therefore shown to recreate the payouts of dynamic trading strategies through passive positions in LP shares. Rather than using dynamic trading to replicate a desired payoff, a user may instead purchase and hold the corresponding LPs, while rebalancing is handled by an external group of arbitrage-seeking traders.

\section{Assumptions and Notation}
\subsection{Geometric Mean Market Makers (G3Ms)} \label{sec:g3ms}

A Geometric Mean Market Maker (G3M) is an Automated Market Maker (AMM) \cite{hanson_lmsr} whose feasible trade set is determined by the weighted geometric mean of its reserves. Specifically, for a set of $n$ assets with corresponding weight vector $w(t)=(w_1(t),\ldots,w_n(t))$ and reserve vector $R(t)=(R_1(t),\ldots,R_n(t))$ with $R(t) \in \mathbb{R}^n_+$, a G3M enforces the geometric mean

\begin{equation}
V(t)=\prod_{i=1}^{n}R_i(t)^{w_i(t)}
\end{equation}
for all $t \geq 0$. By assumption, the weight vector is satisfies
\begin{align} 
\label{eq:ws_sum} \sum_{i=1}^{n}{w_i}(t)=1, \\
w_i(t) \geq 0 \label{eq:ws_geq}.
\end{align}

\noindent A feasible trade is one that results in an updated reserve vector $R'(t)=(R'_1(t),\ldots,R'_n(t))$ for which 
\begin{equation} \nonumber
V(t)=\prod_{i=1}^{n}R'_i(t)^{w_i(t)}.
\end{equation}

\noindent In this paper, we work with G3Ms with no fees. This allows us to greatly simplify the results, while providing a close approximation for many real-world G3Ms that charge traders a small fee. In this setting, let the feasible trades for a G3M be defined as the set of vectors of the form $\Delta(t)=(\Delta_1(t),\ldots,\Delta_n(t))$ with $\Delta(t) \in \mathbb{R}^n_+$ that satisfy

\begin{equation} \nonumber
V(t)=\prod_{i=1}^{n}R_i(t)^{w_i(t)}=\prod_{i=1}^{n}(R_i(t)+\Delta_i(t))^{w_i(t)},
\end{equation}

\noindent with $\Delta_i(t)$ representing the amount of asset $i$ that a trader will deposit into the pool. (Negative values indicate amounts the trader removes from the pool.)

For a given weighted geometric mean, $V(t)$, the price offered by a G3M depends only on the size of the trade and the balances of reserves in the LP. Denote the prices of the assets in the reserve by the vector $S(t)=(S_1(t),\ldots,S_n(t))$ with  $S(t) \in \mathbb{R}^n_+$. As shown in~\cite[Eq.~7]{balancer_wp}, no-arbitrage requires that for all $i \neq j$,

\begin{equation} \label{eq:noarb}
\frac{{R_i(t)}/{w_i(t)}}{{R_j(t)}/{w_j(t)}}=\frac{S_j(t)}{S_i(t)}.
\end{equation}

\noindent That is, if the weight-normalized ratio of reserves for two assets in the LP is equal to the ratio of their prices, then no arbitrage opportunity exists. We denote the payoff of the LP at time $t$ by $G(t)$. Since LP shares can be redeemed at any time for their underlying assets, their payoff is equal to the value of the underlying reserves:

\begin{equation} \label{eq:sum_reserves}
G(t)=\sum_{i=1}^{n} R_i(t)S_i(t).
\end{equation}

\noindent From \eqref{eq:noarb} and \eqref{eq:sum_reserves}, we have for all $j \in (1,\ldots,n)$

\begin{align} \label{eq:tws_rebalancing}
G(t)&=\frac{R_j(t)S_j(t)}{w_j(t)}.
\end{align}

\noindent Note that \eqref{eq:tws_rebalancing} is equivalent to $R_i(t)S_j(t)=w_j(t)G(t)$. In other words, the no-arbitrage condition ensures that the value of the position in asset $i$ represents a proportion $w_i$ of the LP's overall value. As shown in~\cite{AC20} and~\cite{balancer_wp}, should asset values in the LP deviate from the target weights, an arbitrage opportunity is created to restore \eqref{eq:tws_rebalancing}. To preclude arbitrage, the G3M LP is therefore continually rebalanced so that the proportion of value allocated to each asset $j$ matches its target weight, $w_j(t)$, akin to an ETF. Using \eqref{eq:noarb} and \eqref{eq:tws_rebalancing}, and noting the restriction \eqref{eq:ws_sum}, one can derive the LP share payoff (total value of assets it can be redeemed for) as a function of the weighted geometric mean $V(t)$:

\begin{align} \nonumber
G(t) & = \frac{R_j(t)S_j(t)}{w_j(t)}\prod_{1\leq j<i\leq n}\left(\frac{R_i(t)S_i(t)w_j(t)}{R_j(t)S_j(t)w_i(t)}\right)^{w_i(t)} \\ \nonumber
& = \prod_{i=1}^{n}\left(\frac{R_i(t)S_i(t)}{w_i(t)}\right)^{w_i(t)} \\
& = V(t)\prod_{i=1}^{n}\left(\frac{S_i(t)}{w_i(t)}\right)^{w_i(t)}, \label{eq:gen_payoff}
\end{align}

\noindent where in the second step we use $w_j(t) = 1-\sum_{i \neq j} w_i(t)$. In the case where weights are constant, all trades will preserve the weighted geometric mean, so $V(t)=V(0)$ for all $t$. Section 3 makes use of this fact to price G3Ms with constant weights using \eqref{eq:gen_payoff}. When $w(t)$ is a more general adapted process, we must specify the evolution of $V(t)$, which may be a stochastic process. This problem is taken up in Section 4.

\subsection{Market Model} \label{sec:market_model}

Let $(\Omega,\mathcal{F},\{\mathcal{F}_t\}_{t \in \mathbb{R}}, \mathbb{P})$ be a conventional filtered probability space \cite{shreve2004stochastic}. Assume frictionless markets, and consider a financial market that consists of $d$ risky assets and one money market (risk-free) asset. For pricing applications, assume further that there exists an equivalent probability measure $\tilde{\mathbb{P}}$ such that the money market asset and risky assets have respective stochastic differentials

\begin{equation}
dM(t)=M(t)r(t)dt
\end{equation}
and
\begin{equation} \label{eq:risky_assets}
dS_i(t)=S_i(t) \left[r_i(t)dt +\sum_{j=1}^{d} \sigma_{ij}(t)dB_j(t)\right], \qquad i,j \in \{1,\ldots,d\}.
\end{equation}

\noindent Here, $B(t)=(B_1(t),\ldots,B_d(t))$ is a standard Brownian motion under $\tilde{\mathbb{P}}$, $r(t)$ is the riskless interest rate and the components of the volatility matrix, $(\sigma_{ij}(t))_{i=1,\ldots,d;j=1,\ldots,d}$, are adapted processes. Allowing pairwise correlation between risky assets prices, we can rewrite \eqref{eq:risky_assets} as

\begin{equation} 
dS_i(t)=S_i(t)\left[r(t)dt + \sigma_{i}(t)dW_i(t) \right],
\end{equation}

\noindent where each $W_i(t)=\sum_{j=1}^{d}\int_{0}^{t}\frac{\sigma_{ij}(u)}{\sigma_i(u)}dB_j(u)$ is a Brownian motion (by L\'{e}vy's theorem for characterizing a Brownian motion), and $\sigma_i(t)=\sqrt{\sum_{j=1}^d\sigma_{ij}^2(t)}$ is the volatility of asset $i$ which we assume is never zero. Define

\begin{equation} \nonumber
dW_i(t)dW_j(t)=\rho_{ij}(t)dt,
\end{equation}

\noindent where $\rho_{ij}(t)$ is the instantaneous correlation between the Brownian motions $W_i(t)$ and $W_j(t)$. It can be shown that $0 \leq \rho_{ij}(t) \leq 1$.

\section{Constant-Weight G3Ms}

In this section, the prices associated with the payoff in \eqref{eq:gen_payoff} are derived in the case of constant-weight G3Ms. Working with G3Ms consisting of $n \leq d$ risky assets, we use the model of \S\ref{sec:market_model} and assume the volatility matrix $(\sigma_{ij}(t))_{i=1,\ldots,n;j=1,\ldots,d}$ and the interest rate price process $r(t)$ are constant; we set $\sigma_i(t)=\sigma_i \geq 0$ and $r(t)=r$ for all $t$. Note when the weights are fixed, $V(t)$ will be constant. The value of an $n$-asset LP with constant weights $w_i(t)=w_i$ is therefore given by the discounted time-$t$ expectation of \eqref{eq:gen_payoff} under the risk-neutral probability measure, $\tilde{\mathbb{P}}$. Denote the value of the LP share at time $t$ by

\begin{align} \label{eq:pricing_function}
f(t,S(t))&=\tilde{\mathbb{E}}\left[e^{-r(T-t)}G(T) |\mathcal{F}(t) \right],
\end{align}

\noindent where $S(t)$ is the vector of time-$t$ prices for the reserve assets in the LP. This leads to the following Proposition.

\begin{proposition}[Pricing constant-weight LPs] \label{prop:constant-weigh} The price of the LP share with payoff \eqref{eq:gen_payoff} and constant weights $w_i(t)=w_i$ is given by the discounted expectation in \eqref{eq:pricing_function} and is equal to

\begin{align} \label{eq:no_arb_price}
f(t,S(t)) &=
e^{\eta}V(0)\prod_{i=1}^{n}\left(\frac{S_i(t)}{w_i} \right)^{w_i} \\
&=G(t)e^{\eta} \label{eq:no_arb_price_eta},
\end{align}

\noindent where
\begin{align} \label{eq:constant_eta}
\eta=\frac{1}{2}\left(\sum_{i=1}^{n}{\sigma_i^2}(w_i^2-w_i) + \sum_{i \neq j}\sigma_i \sigma_j\rho_{ij}w_iw_j \right) (T-t).
\end{align}

\noindent Furthermore, $\eta \leq 0$.

\end{proposition}

\noindent From the general case addressed in Proposition \ref{prop:constant-weigh}, one can recover the result in~\cite[Appendix C]{ACC19} for the simpler Uniswap constant-product market.

\begin{corollary}[Pricing Uniswap LP shares] \label{prop:uniswap} Define Uniswap as a G3M with $n=2$ assets, $a$ and $b$, and $w_a=w_b=\frac{1}{2}$. Then the Uniswap LP has

\begin{align} \label{eq:uniswap_eta}
\eta_{U}=-\frac{\sigma_{r_{ab}}^2}{8}(T-t).
\end{align}, 

\noindent where

\begin{equation}
\sigma_{r_{ab}}=\sqrt{\sigma_a^2 + \sigma_b^2 - 2\sigma_a\sigma_b\rho_{ab}}.
\end{equation}

\noindent In particular, we prove that $\sigma_{r_{ab}}$ is the volatility of the price ratio $S_a/S_b$ for the two assets in the LP.

\end{corollary}

\noindent \textbf{Volatility Losses.} To understand the content of $\eta$ in \eqref{eq:no_arb_price_eta} and \eqref{eq:constant_eta}, recall the observation in \S\ref{sec:g3ms} that no-arbitrage requires the G3M LP to continually rebalance its reserves to match the target weights. Should asset values in the LP deviate from the target weights, an arbitrage opportunity is created to restore \eqref{eq:tws_rebalancing}. By definition, arbitrage results in a greater value of assets exiting the LP than entering, which reduces the value of the LP shares. LP shares therefore incur rebalancing costs due to arbitrage in order to enforce a target portfolio composition. To understand the magnitude of these costs, contrast the LP share payoff in this situation with that resulting from continually rebalancing a portfolio to fixed weights under zero transaction costs. From~\cite{bertrand}, the stochastic differential for this portfolio is given by

\begin{equation} \label{eq:constant_mix_payoff}
dN(t)=N(t)\sum_{i=1}^{n}w_i\frac{dS_i}{S_i}.
\end{equation}

\noindent In Appendix \ref{sec:constant_mix}, we show that this portfolio strategy has value

\begin{equation} \label{eq:cm_payoff_eta}
\tilde{\mathbb{E}}\left[e^{-r(T-t)}N(T)| \mathcal{F}(t) \right] = e^{-\eta(T-t)}f(t,S(t)).
\end{equation}

\noindent This shows that $e^{\eta(T-t)}$ represents the expected loss LP shares incur relative to a constant-mix portfolio with equivalent weights. This coincides with the well-documented result of volatility harvesting~\cite{vol_harvest} which states that a continuously-rebalanced constant-mix portfolio has a greater growth rate than the weighted average of its component assets. The constant-mix portfolio in \eqref{eq:cm_payoff_eta} benefits from volatility through the $e^{-\eta(T-t)}$ term, while the no-fee LP in \eqref{eq:gen_payoff} does not. This is the cause of the supermartingale behavior observed in \eqref{eq:no_arb_price_eta}. One can therefore replicate the value of a fixed-weight G3M LP share with less initial capital by continuously rebalancing to the same target weights in a frictionless market. Informally, this occurs because the G3M lags the market during rebalancing. LP rebalancing occurs through arbitrage which results when LP reserves do not reflect updated market prices. LP shares therefore rebalance at suboptimal prices relative to conventional constant-mix portfolios. 

\begin{figure}
   \centering
    \includegraphics[scale=0.45]{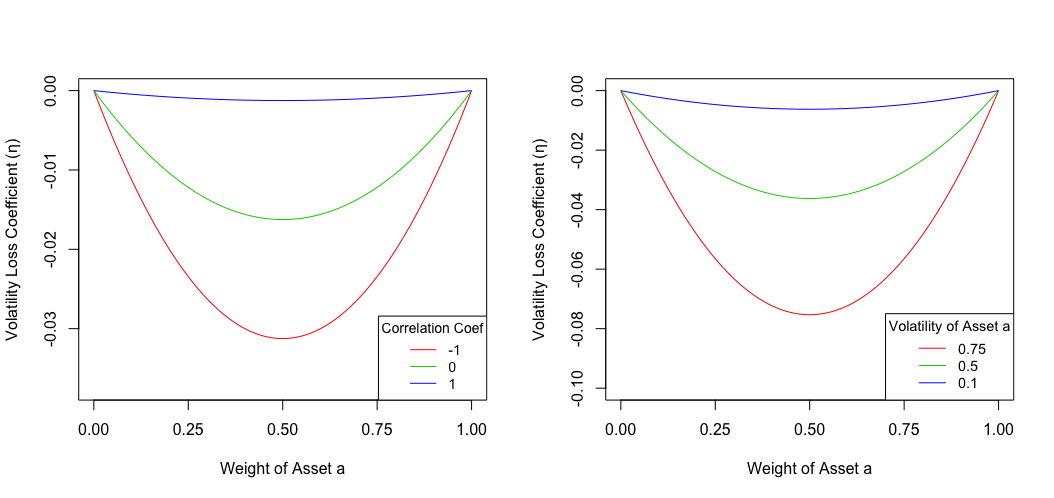}
    \caption{The left figure plots $\eta$ (defined in \eqref{eq:constant_eta}) for a two-asset LP share ($n=2$) with asset volatilities $\sigma_a=0.3$ and $\sigma_b=0.2$, given different choices for the weight $w_a$ of asset $a$ and for the correlation coefficient $\rho_{ab}$. The right figure holds $\rho_{ab}=0$ and plots $\eta$ for different choices of $w_a$ and volatility levels $\sigma_a$.}
    \label{fig:eta_graph}
\end{figure}

Figure \ref{fig:eta_graph} plots $\eta$ using an example of a two-asset LP share with assets $a$ and $b$. Note that \eqref{eq:uniswap_eta} is minimized in the Uniswap configuration, where $w_a=\frac{1}{2}$; this represents the maximum loss relative to the constant-mix portfolio. Meanwhile, $\eta$ is zero when $w_a=0$ and when $w_a=1$; in these cases the LP shares coincide with buy-and-hold portfolios, and there are no opportunities for trading against the assets of the pool (hence no arbitrage losses). The quantity $\eta$ is increasing with respect to the correlation coefficient $\rho_{ab}$. The higher the correlation coefficient, the smaller the price deviations are expected
to be for the assets in the LP; thus, high values of $\rho_{ab}$ limit arbitrage losses. Similarly, higher levels of volatility for one of the two assets in the LP produce greater volatility losses. In the case two-asset case, when $\sigma_a=\sigma_b$ and $\rho_{ab}=1$, $\eta$ is zero regardless of the choice of weight, as there is no expected trading (price moves are expected to have identical magnitude and direction).

\paragraph{LP share gamma.} Taking the first derivative of \eqref{eq:no_arb_price} with respect to the stock price (``delta" in options terminology) yields $f_{S_i} = w_iS_i^{-1}f$, which is non-negative. Taking the second derivative (``gamma") gives $f_{S_iS_i} = w_i(w_i-1)S_i^{-2}f$, which, by the restrictions on $w_i$, is non-positive. The constant-weight LP will therefore decrease its unit position in asset $i$ as its price increases (and conversely increase its unit position as price declines). The resulting payoff is concave in $S_i$, an effect Uniswap traders refer to as ``impermanent loss." Specifically, regardless of the direction of a price movement, the LP share will decrease in value relative to the buy-and-hold portfolio, which has a gamma of zero. Note that the constant-mix portfolio without rebalancing costs described above also exhibits a negative gamma, but, unlike the G3M LP share, it benefits from volatility in exchange (this is the content of \eqref{eq:cm_payoff_eta}). The LP share's gamma is minimized (impermanent loss is highest) when the weight of asset $i$ is $w_i=\frac{1}{2}$, while it is zero when $w_i=1$ (LP holds only asset $i$) and when $w_i=0$ (no exposure to $i$). As noted in~\cite{8020pools}, this comes with a direct trade-off to the slippage offered to traders in the pool.

\section{General Weight Functions}

\subsection{Discrete-Time Weighted Geometric Mean}

In this section, the payoffs for G3M LP shares are derived for the case where the weight vector $w(t)$ is an $\mathcal{F}(t)$-measurable process. From an initial weighted geometric mean $V(0)$, assume the process $V(t)$ is generated by updating the weight vector at a sequence of re-weighting times $0=t_0<t_1<\ldots<t_s=T$. The weight vector is updated at the left endpoint of each interval $[t_k,t_{k+1})$ and is then held constant until the next re-weighting time. This ensures that $V(t)$ remains constant on each interval but is allowed to vary across intervals. Assume the initial weighted geometric mean $V(0)=V(t_0)$ is given by

\begin{equation} \nonumber
V(t_0)=\prod_{i=1}^{n}R_i(0)^{w_i(0)}.
\end{equation}

\noindent By assumption, updating satisfies $V(t_{k-1})=\prod_{i=1}^{n}R_i(t_k)^{w_i(t_{k-1})}$ and $V(t_k)=\prod_{i=1}^{n}R_i(t_k)^{w_i(t_k)}$. Since the weighted geometric mean is constant within each interval, at each $t_k$ we have

\begin{align} \nonumber
V(t_k)&=\prod_{i=1}^{n}R_i(t_k)^{w_i(t_k)} =\prod_{i=1}^{n}R_i(t_{k})^{w_i(t_{k-1})}R_i(t_{k})^{\Delta w_i(t_k)} =V(t_{k-1})\prod_{i=1}^{n}R_i(t_{k})^{\Delta w_i(t_k)},
\end{align}

\noindent where $\Delta w_i(t_k)=w_i(t_k)-w_i(t_{k-1})$ and $\sum_{i=1}^{n} \Delta w_i(t)=0$. Repeating this procedure starting from $t_s$ we get

\begin{equation} \nonumber
V(t_s)=V(0)\prod_{k=1}^{s} \prod_{i=1}^{n}R_i(t_{k-1})^{\Delta w_i(t_k)}. 
\end{equation}

\noindent Solving for $R_i(t_{k-1})$ in the no-arbitrage condition of \eqref{eq:tws_rebalancing}, we have

\begin{equation} \nonumber
R_i(t_{k-1})=\frac{w_i(t_{k-1})}{S_i(t_{k-1})}G(t_{k-1}).
\end{equation}

\noindent Again using  $\sum_{i=1}^{n} \Delta w_i(t)=0$,

\begin{align} \nonumber
\prod_{i=1}^{n}R_i(t_{k-1})^{\Delta w_i(t_k)} = \prod_{i=1}^{n}\left(\frac{w_i(t_{k-1})}{S_i(t_{k-1})}\right)^{\Delta w_i(t_k)}. 
\end{align}

\noindent This provides the discrete-time formula for the weighted geometric mean at time $T$:

\begin{equation} \label{eq:discrete_payoff}
V(T)=V(t_0) \prod_{i=1}^{n}\prod_{k=1}^{s}\left(\frac{w_i(t_{k-1})}{S_i(t_{k-1})}\right)^{\Delta w_i(t_k)}.
\end{equation}

\noindent Note that this discrete-time formulation is the most realistic setting for G3Ms deployed on public blockchains such as Ethereum that have positive-length time intervals between blocks. In this setting, each weight adjustment will present an arbitrage opportunity that results in some value loss for LP shares. 

\subsection{Payoff for Continuously-Varying Weights}

This section studies LP returns in the case where weights are allowed to vary continuously. The key result of this section is the following.

\begin{proposition}[Payoff for dynamic-weight LPs] \label{prop:dynamic_weights}
Assume each component weight function $w_i(s)$, $i \in \{1,\ldots,n\}$, is continuous and has bounded variation, and denote the length of the longest interval in \eqref{eq:discrete_payoff} by $||\Pi||=\max_{k=0,\ldots,s-1}(t_{k+1}-t_k)$. Then taking the limit in \eqref{eq:discrete_payoff} as $||\Pi||\to 0$ gives the weighted geometric mean for all $T \geq t \geq 0$

\begin{align} \nonumber
V(T) = V(t)\prod_{i=1}^{n}\left(\frac{w_i(T)}{S_i(T)}\right)^{w_i(T)}\left(\frac{S_i(t)}{w_i(t)}\right)^{w_i(t)}e^{\int_{t}^{T}w_i(t)d\log(S_i(t))}
\end{align}

\noindent with corresponding payoff function

\begin{align} \label{eq:gen_dynamic_payoff}
G(T)=G(t)\prod_{i=1}^{n}e^{\int_{t}^{T}w_i(t)d\log(S_i(t))}.
\end{align}

\end{proposition}

\noindent This is the payoff function we work with in the remaining sections. \\

LP prices computed by taking discounted risk-neutral expectations in \eqref{eq:gen_dynamic_payoff} will depend on the stochastic process chosen for the weight vector $w(t) = (w_1(t),\ldots,w_n(t))$. However, if the weight vector is a deterministic function of time, the solution can be simplified. In this case, LP prices can be computed directly given the model in \S\ref{sec:market_model}.

\begin{proposition}[Pricing LPs with deterministic time-varying weights] \label{prop:time_weights} If each component of $w(t) = (w_1(t),\ldots,w_n(t))$ is an $\mathcal{F}(t)$-measurable deterministic function of $t$, then the corresponding LP share price is given by the discounted expectation under the risk-neutral measure of \eqref{eq:gen_dynamic_payoff} and is equal to

\begin{align}
\tilde{\mathbb{E}}\left[e^{-r(T-t)}G(T)| \mathcal{F}(t) \right] &= G(t)e^{\eta(t,T)},
\end{align}

\noindent where

\begin{align} \label{eq:eta_time}
\eta(t,T) = \sum_{i=1}^{n}\frac{\sigma_i^2}{2}\int_{t}^{T}[w_i^2(t)-w_i(t)]dt + \frac{1}{2}\sum_{i \neq j}\sigma_i \sigma_j\rho_{ij} \int_{t}^{T}w_i(t)w_j(t)dt. 
\end{align}

\end{proposition}

\noindent These prices are relevant to applications that require G3M weights to be adjusted according to a fixed schedule. Typically, an LP will reduce the weight of one of its assets until some target weight is reached. This creates an arbitrage opportunity to remove units of the asset whose weight is declining in favor of the other reserve assets. This has been proposed as a mechanism for bootstrapping liquidity in nascent markets~\cite{lbps}. Similarly, it may be desirable for an LP to decrease its exposure to assets with fixed maturities, such as options and bonds, as these near expiry.

\section{Payoff Targeting and Replication}

This section shows how to select G3M weight functions to ensure that the resulting payoffs of the LP shares replicate the payoffs of derivative claims on the price of an asset. We work with a two-asset G3M that consists of a risky asset with weight $w(x,t)$ and a position in the risk-free asset with weight $1-w(x,t)$, where $x=S_\alpha(t)$ is the price of the risky asset. Consider a contract with payoff given by the real-valued function $g(x,t)$.\footnote{ For example, a forward contract expiring at time $T$ has $g(x,T)=S_\alpha(T)-K$, and an option expiring at time $T$ has $g(x,T)=\max\{S_\alpha(T)-K,0\}$, where, in both examples, $K$ is the strike price.} Rewriting \eqref{eq:gen_dynamic_payoff} as

\begin{align} \label{eq:2var_payoff}
G(t)=G(0)e^{\int_{0}^{t}w(x,s)d\log(x)},
\end{align}

\noindent we solve for the weight $w^{*}(x,t)$ such that the LP and the derivative contract have the same payoff for all $t \geq 0$:\footnote{In practice, enforcing weight updates of this form may require the use of a ``price oracle" such as~\cite{uma} that reports the price of the asset to the G3M smart contract.}

\begin{equation} \label{eq:target}
    G(t) = g(S_\alpha(t),t) \quad \text{for all }  t \geq 0.
\end{equation}

\begin{proposition}[Replicating weight function] \label{prop:target_weight} Let $g$ be differentiable with respect to $x$ for $x \in \mathbb{R}_+$. Then the solution for $w(x,t)$ in \eqref{eq:target} with initial condition $G(0) = g(S_\alpha(0),0)$ is given by

\begin{align} \label{eq:w_star}
w^{*}(x,t) = \frac{d \log(g(x,t)))}{d \log(x)}=  \frac{xg_x(x,t)}{g(x,t)},
\end{align}

\noindent where $g_x$ is the partial derivative of $g$ with respect to $x$. The payoff $g(x,t)$ can be replicated by a G3M LP provided that $w^{*}(x,t)$ is continuous in $x$ and

\begin{equation}  \label{eq:w_condition}
    0 \leq  w^{*}(x,t) \leq 1 \quad \text{for all } x,t \in \mathbb{R}_+.
\end{equation}

\end{proposition}

\noindent Equation \eqref{eq:w_star} is the elasticity of a contingent claim, i.e.~the percent change in the value of the derivative given a one-percent change in the price of the risky asset (it is also termed ``lambda" or ``omega" in derivatives parlance). The condition in \eqref{eq:w_condition} is due to the restrictions \eqref{eq:ws_sum} and \eqref{eq:ws_geq} on the weights of the G3M. Note that if short-selling an LP share is possible,  one can also replicate claims with $-1 \leq  w^{*}(x) \leq 0$. The condition \eqref{eq:w_condition} states that the G3M cannot be used to gain leverage on its reserve assets. The maximum elasticity of a contingent claim with respect to the risky asset is therefore attained when $w(x,t)=1,$ when the pool consists exclusively of the risky asset. 
For differentiable claims where \eqref{eq:w_condition} is satisfied, \eqref{eq:w_star} guarantees that holding an LP share provides an exact static hedge of the contingent claim regardless of the model one uses for the underlying asset price. In practice, continuous weight adjustments will not be possible in the discrete-time setting of public blockchains. Discrete weight adjustments will result in arbitrage opportunities that reduce the value of the pool. This implies that the LP share will in practice provide a sub-hedge for $g(x,t)$, though the introduction of fees can be used to offset all or part of these relative losses. 

It will often be possible to relax the assumptions of Proposition \ref{prop:target_weight} by instead replicating the \emph{value} of the contract by replacing $g(x,t)$ in \eqref{eq:w_star} with its discounted expectation under the risk-neutral measure. Such pricing formulae will typically require the use of a model such as that of \S\ref{sec:market_model} for the underlying price. The resulting LP share will provide a parametric hedge for the derivative asset, and the accuracy of the hedge will depend on the model chosen. For concreteness, we provide an example below. \\

\textbf{Example} (Protective put). A protective put~\cite{McKeon65} is a popular risk-management strategy wherein an investor buys an asset alongside a put option on the same asset. In exchange for the option premium, the strategy allows the investor to profit from price appreciation while being protected from losses. Given a model for the option price, we can show that a G3M LP can be programmed to synthetically replicate a protective put. For example, using the Black--Scholes formula~\cite{black_scholes} for the value of a put option, we have 

\begin{equation} \nonumber
   P(x,t) = Ke^{-r(T-t)}\Phi(-d_2)-x\Phi(-d_1),
\end{equation}

\noindent where $T>0$ is the expiration, $K \geq 0$ is the strike price, $\Phi(\cdot)$ is the standard normal CDF, and

\begin{align} \nonumber
   & d_1 = \frac{\log(x/K)+ (r + \sigma_\alpha^2/2)(T-t)}{\sigma_\alpha \sqrt{T-t}}, \\
   & d_2 = d_1 - \sigma_\alpha \sqrt{T-t}. \nonumber
\end{align}

\noindent where $\sigma_\alpha$ is the volatility of the risky asset. It can be shown that the protective put claim $g(x,t) = x + P(x,t)$ has elasticity

\begin{align} \label{eq:protput_weight}
    w^{*}_{pp}(x,t) &= \frac{x(1 - \Phi(-d_1))}{P(x,t)+x}.
\end{align}

\noindent Note that the numerator is equal to the price of the asset multiplied by one plus the ``put delta," the first derivative of the put with respect to $x$. This quantity is always non-negative, as $0 \leq \Phi(\cdot) \leq 1$ and $x \in \mathbb{R}_+ $. The denominator is also non=negative, as the value of the option is given by the time-$t$ risk-neutral expectation of $g(x,T)=\max\{S_\alpha(T)-K, 0\}$. Therefore $w^{*}_{pp}(x,t) \geq 0$. Furthermore,

\begin{align} \nonumber
   w^{*}_{pp}(x,t) \leq w^{*}_{pp}(x,t) +  \frac{ Ke^{-r(T-t)}\Phi(-d_2)}{P(x,t)+x} = 1.
\end{align}

\begin{figure}
    \includegraphics[scale=0.45]{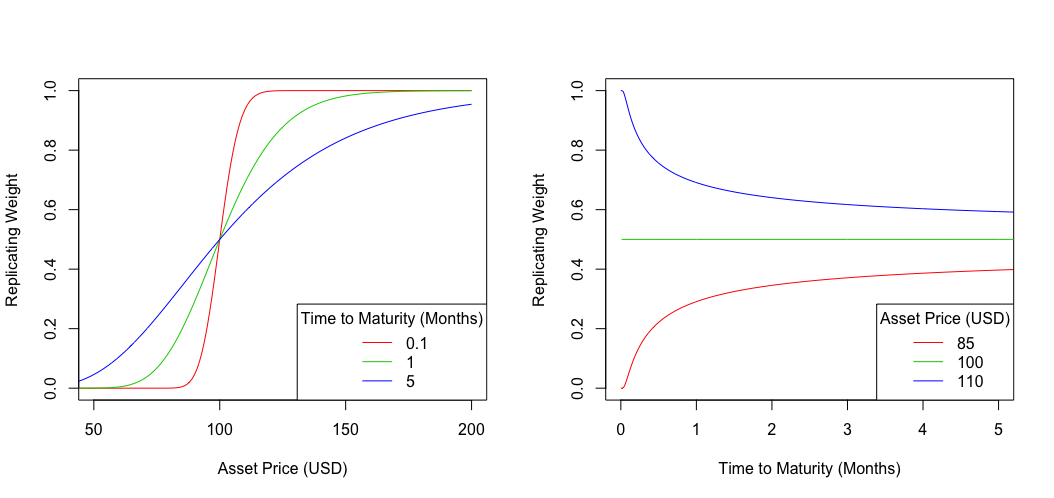}
    \caption{Replicating weights $w(x,t)$ for a protective put given a strike price of $K=100$ USD for the put option and a risky asset with monthly volatility of $\sigma_{\alpha} =0.2$. The left figure plots the replicating weight as a function of asset price for different maturities, while the right figure plots the replicating weight as a function of time to maturity for different values of the asset price.}
    \label{fig:replicating_weights}
\end{figure}

\noindent We conclude that setting the G3M's weight for the risky asset to \eqref{eq:protput_weight} replicates a protective put on the risky asset with strike $K$ and expiry $T$. Using the same procedure, we can show that an LP can replicate a covered call, which consists of a long position in an asset alongside a short position in a call option written on the same asset. Figure \ref{fig:replicating_weights} shows the weight function that replicates a protective put. As the price of the underlying asset increases, the weight tends to one, where the LP consists entirely of the risky asset. As price declines, the LP increases the weight of the money market (risk-free) asset. The relationship with time to maturity depends on whether the put option is ``in the money" (above the strike price $K$). If the put is ``at the money" ($S_\alpha=K$), then the G3M weight is $0.5$ regardless of time to maturity. If the put is near expiry and $S_\alpha>K$, then the G3M places a greater weight on the risky asset. If the put is near expiry and $S_\alpha<K$, then the G3M places a greater weight on the risk-free asset. The replicating weight of the protective put in the risky asset is therefore increasing with respect to the probability that the put will expire out of the money. \\

A number of interesting derivative contracts, such as pure (``naked") options, often exhibit elasticity far greater than one. There are two approaches to replicating such contracts. The first involves taking offsetting positions in addition to the LP. For example, holding an LP that replicates a protective put while also establishing a short position in the underlying asset will replicate the payoff of the put option. Using the approach of the proceeding example, it can be shown that a portfolio consisting of a call option plus a position worth $e^{r(T-t)}K$ in the money market satisfies \eqref{eq:w_condition}. Holding the replicating LP share of this portfolio in addition to an offseting short position of $e^{r(T-t)}K$ in the money market will replicate the pure call option. The offsetting positions in the risky or money market assets can be interpreted as borrowing the respective assets and placing them in the replicating LP. This could be facilitated by an existing lending protocol such as \cite{aave_lps} that accepts LP shares as collateral for secured loans. For example, to replicate a naked put option, the investor would place an amount of capital equal to the initial price of the option in a G3M that replicates a protective put. At the same time, the lending protocol would supply one unit of the risky asset to the G3M, while taking the corresponding LP shares as collateral. Even if the option expires worthless, the lender can be assured that the replicating LP will be at least as valuable as the risky asset that was lent, ensuring repayment of the loan. At expiration, after repaying the borrowed asset to the lending protocol, the investor's remaining position will have equal value to that of the pure put option (assuming the model used in constructing the hedge was correctly calibrated).

A second approach to replicating claims with elasticity greater than one involves adding derivatives to a G3M's reserves. The use of levered assets can expand the range of derivatives that an LP share can be used to replicate. For example, in place of the risky asset one can include a derivative on the risky asset in the LP's reserves with time-$t$ price $z(S_\alpha(t))$. In this case \eqref{eq:2var_payoff} becomes

\begin{align} \label{eq:payoff_deriv}
G(t)=G(0)e^{\int_{0}^{t}w(z(x))d\log(z(x))},
\end{align}

\noindent and we have the following solution. For simplicity, we work with the single-variable payoff, $g(x)$.

\begin{corollary}[Replication with derivative assets] \label{prop:deriv_target} Let $g$ and $z$ be differentiable on $\mathbb{R}_+$. Then the solution to $G(t) = g(S_\alpha(t))$ when $G(t)$ is given by \eqref{eq:payoff_deriv} and with initial condition $G(0) = g(S_\alpha(0))$ is

\begin{align}
& w^{*}(z(x)) = \frac{d \log(g(x))}{d \log(z(x))}.
\end{align}

\noindent Replication with a G3M LP requires that

\begin{align} 
 0 \leq \frac{d \log(g(x))}{d \log(z(x))} \leq 1 \quad \text{for all } x \in \mathbb{R}_+. 
\end{align}

\end{corollary}

\noindent G3Ms can therefore replicate any claim whose logarithmic derivative is no larger than that of its reserve asset price function. The logarithmic derivatives of the payoff $g(x)$ and price $z(x)$ determine their infinitesimal relative changes and can informally be thought of as a measure of leverage. When the target claim is no more levered than the reserve claim, replication will be possible through a static position in the LP.

\section{Conclusion}

This work studies the returns investors receive for contributing reserves to G3Ms. We derive explicit payoff and pricing functions for LP shares in G3Ms that utilize both static and dynamic weights. We show that LP share payoffs of G3Ms that do not charge fees are supermartingales under the risk-neutral probability measure, due to having higher rebalancing costs than constant-mix portfolios. Utilizing dynamic weights, we show that G3M LP shares can be used to provide exact static hedges for arbitrary financial contracts whose payoffs have elasticity between zero and one. In a parametric setting, we demonstrate how to use offsetting positions and external leverage to replicate more general financial contracts, such as standard options. 

A question left open by this paper concerns fees. In practice, most G3Ms charge fees that introduce path dependencies in LP share payoffs~\cite{AC20}. As fees may alter both the frequency and the cost of G3M rebalancing, it may be instructive to consider the corresponding constant-mix portfolio under rebalancing restrictions and transaction costs~\cite{RePEc:ipg:wpaper:2014-303}. 

\section*{Acknowledgement}

The author would like to thank Guillermo Angeris, Tarun Chitra, Alexandre Obadia and Assimakis Kattis for their feedback on this paper.

\clearpage

\appendix
\section{Proofs}

\subsection{Combining Brownian Motions} \label{sec:lemma_bms}

We establish a definition that will be useful in the proofs of Propositions \ref{prop:constant-weigh} and \ref{prop:time_weights}. For $n \leq d$, and given that the components of $w(t)$ are square-integrable by the restrictions in \eqref{eq:ws_sum} and \eqref{eq:ws_geq}, we can define

\begin{equation} \nonumber
Z_P(t)=\sum_{i=1}^{n}\int_{0}^{t}\frac{w_i(u)\sigma_{i}(u)}{\sigma_P(u)}dW_j(u),
\end{equation}

\noindent with

\begin{align} \nonumber
\sigma_P(t)=\sqrt{\sum_{i=1}^{n}w_i^2(t)\sigma_i^2(t) + \sum_{i \neq j} w_i(t) w_j(t)\sigma_i(t)\sigma_j(t)\rho_{ij}(t)},
\end{align}

\noindent which we assume is non-zero. (As will be discussed in the proofs of Propositions \ref{prop:constant-weigh} and \ref{prop:time_weights}, $\sigma_P$ represents the volatility of the weighted geometric mean of the risky asset prices.) We can use these definitions to write

\begin{equation} \nonumber
\sigma_P(t)dZ_P(t) = \sum_{i=1}^{n}w_i(t)\sigma_i(t)dW_i(t).
\end{equation}

\noindent It is trivial to verify that $Z_P$ has quadratic variation $\langle Z_P(t) \rangle = t$. Being the sum of continuous martingales, $Z_P(t)$ is therefore a Brownian motion by L\'{e}vy's theorem.

\subsection{Proof of Proposition \ref{prop:constant-weigh}}

\noindent The proof of Proposition \ref{prop:constant-weigh} has two parts: first we prove \eqref{eq:no_arb_price_eta}, and then we prove that the quantity $\eta$ defined in \eqref{eq:constant_eta} is at most zero.  

\paragraph{i)} The proof of \eqref{eq:no_arb_price_eta} runs as follows: the differential for the weighted geometric mean of the prices will give a geometric Brownian motion, from which \eqref{eq:no_arb_price_eta} follows immediately by taking expectations in \eqref{eq:pricing_function}. 

Note that $S_i(t)^{w_i}$ is given by

\begin{align} \nonumber
    S_i^{w_i}(t)=S_i^{w_i}(0)e^{w_i(r-\sigma_i^2/2)t+w_i\sigma_i W_i(t)}.
\end{align}

\noindent Applying It\^o's lemma results in the differential

\begin{align} \nonumber
   dS_i^{w_i}(t)=S_i^{w_i}(t)\left[(w_ir+ \frac{\sigma_i^2}{2}(w_i^2-w_i))dt+w_i \sigma_i dW_i(t)\right],
\end{align}

\noindent which defines a geometric Brownian motion with mean $(w_ir+ \frac{\sigma^2}{2}(w_i^2-w_i))$ and volatility $w_i \sigma_i$. Note further that

\begin{align} \nonumber
\nonumber d(S^{w_i}_i(t)S^{w_j}_j(t)) &= S^{w_i}_i(t)dS^{w_j}_j(t) + dS^{w_i}_i(t)S^{w_j}_j(t) + dS^{w_i}_i(t)dS^{w_j}_j(t) \\
&= S^{w_i}_i(t)S^{w_j}_j(t)[(r(w_i+w_j)+\frac{\sigma_i^2}{2}(w_i^2-w_i)+\frac{\sigma_j^2}{2}(w_j^2-w_j) + w_iw_j\sigma_i\sigma_j\rho_{ij})dt \nonumber \\ &+  w_i\sigma_idW_i(t) + w_j\sigma_jdW_j(t)]. \nonumber
\end{align}

\noindent Iterating gives 

\begin{align} \label{eq:wgm_prices}
d\left(\prod_{i=1}^{n}S^{w_i}_i(t)\right)
&= \prod_{i=1}^{n}S^{w_i}_i(t)\left[\left(r+\sum_{i=1}^{n}\frac{\sigma_i^2}{2}(w_i^2-w_i)+\frac{1}{2}\sum_{i \neq j} w_iw_j\sigma_i\sigma_j\rho_{ij} \right)dt + \sum_{i=1}^{n} w_i\sigma_idW_i(t) \right].
\end{align}

\noindent As shown shown in \S\ref{sec:lemma_bms}, we may define

\begin{align} \nonumber
\sigma_P=\sqrt{\sum_{i=1}^{n}w_i^2\sigma_i^2 + \sum_{i \neq j} w_i w_j\sigma_i\sigma_j\rho_{ij}}
\end{align}

\noindent and

\begin{equation} \nonumber
Z_P(t)=\sum_{i=1}^{n}\int_{0}^{t}\frac{w_i\sigma_{i}}{\sigma_P}dW_j(u),
\end{equation}

\noindent which is a Brownian motion. We can then rewrite \eqref{eq:wgm_prices} as

\begin{align} \label{eq:wgm_prices_result}
d\left(\prod_{i=1}^{n}S^{w_i}_i(t)\right)
&= \prod_{i=1}^{n}S^{w_i}_i(t)\left[\left(r+\sum_{i=1}^{n}\frac{\sigma_i^2}{2}(w_i^2-w_i)+\frac{1}{2}\sum_{i \neq j} w_iw_j\sigma_i\sigma_j\rho_{ij} \right)dt + \sigma_PdZ_P(t) \right], 
\end{align}

\noindent which is a geometric Brownian motion with mean $r+\sum_{i=1}^{n}\frac{\sigma_i^2}{2}(w_i^2-w_i)+\frac{1}{2}\sum_{i \neq j} w_iw_j\sigma_i\sigma_j\rho_{ij}$ and volatility $\sigma_P$. We obtain the result \eqref{eq:no_arb_price_eta}  by taking the expectation in \eqref{eq:pricing_function}. The result in \eqref{eq:no_arb_price} follows from noting that $V(0) = G(0)\prod_{i=1}^{n}\left(\frac{w_i(0)}{S_i(0)}\right)^{w_i(0)}$, which follows from \eqref{eq:gen_payoff}.

\paragraph{ii)} Next, we show that $\eta \leq 0$ (where $\eta$ is defined in \eqref{eq:constant_eta}). Since $\frac{1}{2}(T-t) \geq 0$, this is equivalent to showing that

\begin{align} \nonumber
\sum_{i=1}^{n}{\sigma_i^2}(w_i^2-w_i) + \sum_{i \neq j}\sigma_i \sigma_j\rho_{ij}w_iw_j \leq 0.
\end{align}

\noindent Recall the restrictions \eqref{eq:ws_sum} and \eqref{eq:ws_geq} and the assumption that $\sigma_1,\ldots,\sigma_n$ are positive constants. Since the second summand is positive and $0 \leq \rho_{ij}(t) \leq 1$, it suffices to show that

$$
\sum^n_{i=1}\sigma^2_i(w^2_i-w_i)+\sum_{i\ne j}\sigma_i\sigma_jw_iw_j\leq 0.
$$

\noindent The left-hand side can be rewritten as

\begin{align*} \nonumber
\sum^n_{i=1}\sigma^2_i(w^2_i-w_i)+\sum_{i\ne j}\sigma_i\sigma_jw_iw_j&=-\sum^n_{i=1}\sigma^2_iw_i(1-w_i)+\sum_{i\ne j}\sigma_i\sigma_jw_iw_j
\\ \nonumber
&=-\sum^n_{i=1}\sigma^2_iw_i\left(\sum^{i-1}_{j=1}w_j+\sum^n_{j=i+1}w_j\right)+2\sum_{1\leq i<j\leq n}\sigma_i\sigma_jw_iw_j
\\ \nonumber
&=-\sum_{1\leq j<i\leq n}\sigma^2_iw_iw_j-\sum_{1\leq i<j\leq n}\sigma^2_iw_iw_j+2\sum_{1\leq i<j\leq n}\sigma_i\sigma_jw_iw_j,
\end{align*}

\noindent where the second line follows from \eqref{eq:ws_sum}. Relabeling indices in the first sum gives

$$
-\sum_{1\leq i<j\leq n}\left(\sigma^2_j+\sigma^2_i-2\sigma_i\sigma_j\right)w_iw_j=-\sum_{1\leq i<j\leq n}\left(\sigma_j-\sigma_i\right)^2w_iw_j\leq 0,
$$

\noindent as desired.

\subsection{Proof of Corollary \ref{prop:uniswap}}

The volatility $\sigma_{ab}$ will follow from the expression for the price ratio. The stochastic differential for the ratio of the prices of two assets $S_{r_{ab}}(t)=S_a(t)/S_b(t)$ is given by

\begin{align} \nonumber
S_{r_{ab}}(t) &= (1/S_b(t))dS_a(t) - (S_a(t)/S_b^2(t))dS_b(t) - (1/S_b^2(t))dS_a(t)dS_b(t) + (S_a(t)/S_b^3(t))(dS_b)^2
\\ &=S_{r_{ab}}(\sigma_b^2(t) - \sigma_a(t)\sigma_b(t)\rho_{ab}(t))dt + S_{r_{ab}}\sigma_{r_{ab}}(t)dZ_r(t), \nonumber
\end{align}

\noindent where 

\begin{equation} \nonumber
\sigma_{r_{ab}}(t)=\sqrt{\sigma_a^2(t) + \sigma_b^2(t) - 2\sigma_a(t)\sigma_b(t)\rho_{ab}(t)}
\end{equation}

\noindent and 

\begin{equation} \nonumber
Z_r(t)=\frac{1}{\sigma_{r_{ab}}}\left(\int_{0}^{t} \sigma_{a}(u)dW_a(u)-\int_{0}^{t}\sigma_b(u)W_b(u)\right);
\end{equation}

\noindent note that $Z_r(t)$ is a Brownian motion. Therefore $S_{r_{ab}}$ is a geometric Brownian motion with drift $\sigma_b^2(t) - \sigma_a(t)\sigma_b(t)\rho_{ab}(t)$ and volatility $\sigma_{r_{ab}}(t)$. Assuming constant volatilities and taking $n=2$ and $w_a=w_b=\frac{1}{2}$ in \eqref{eq:no_arb_price_eta}, we have

\begin{align} \nonumber
\eta=\left(-\frac{\sigma_a^2}{8} - \frac{\sigma_a^2}{8} + \frac{1}{4}\sigma_a \sigma_b\rho_{ab} \right) (T-t) = \frac{\sigma_{r_{ab}}^2}{8}(T-t),
\end{align}

\noindent as desired.

\subsection{Payoff of Constant-Mix Portfolio} \label{sec:constant_mix}

\noindent From \eqref{eq:constant_mix_payoff} we have

\begin{align} \nonumber
dN(t)&=N(t)(rdt + \sum_{i=1}^{n}w_i\sigma_idW_i) \\
&= N(t)(rdt + \sigma_PdZ_P), \label{eq:constant_mix_p}
\end{align}

\noindent which gives

\begin{align} \nonumber
N(t)&=N(0)e^{(r-\frac{\sigma_P^2}{2})t+\sigma_PZ_P(t)}.
\end{align}

\noindent Comparing \eqref{eq:constant_mix_p} with \eqref{eq:wgm_prices_result} shows that the difference between their drift terms is \eqref{eq:constant_eta}, from which the result in \eqref{eq:cm_payoff_eta} follows by taking expectations.

\subsection{Proof of Proposition \ref{prop:dynamic_weights}}

\noindent Take the limit as the quantity  $||\Pi||=\max_{k=0,\ldots,s-1}(t_{k+1}-t_k)$ (the size of the longest time interval in \eqref{eq:discrete_payoff}) tends to zero:
$$
V(T)=V(t_0) \lim_{||\Pi||\to 0}\prod_{k=1}^{s}\prod_{i=1}^{n}\frac{w_i(t_{k-1})}{S_i(t_{k-1})}^{\Delta w_i(t_k)}.
$$
We have
\begin{align} \nonumber
\log[{V(T)/V(t_0)}]&= \lim_{||\Pi||\to 0}\sum_{i=1}^{n}\sum_{k=1}^{s}\log\left(\frac{w_i(t_{k-1})}{S_i(t_{k-1})}\right){\Delta w_i(t_k)} \\ \nonumber
&= \sum_{i=1}^{n}\int_{t_0}^{T}\log\left(\frac{w_i(t)}{S_i(t)}\right)d w_i(t) \\ \nonumber
&= \sum_{i=1}^{n}\left[\log\left( \frac{w_i(T)^{w_i(T)}}{w_i(0)^{w_i(0)}} \right) + w_i(T)-w_i(0) - \int_{t_0}^{T}\log(S_i(t))dw_i(t) \right].
\end{align}

\noindent Note that $\sum_{i=1}^{n}[w_i(T)-w_i(t_0)]=0$, and integrate by parts:

\begin{align} \nonumber
\log[{V(T)/V(t_0)}]=\sum_{i=1}^{n}\left[\log\left(\frac{w_i(T)^{w_i(T)}}{w_i(t_0)^{w_i(t_0)}}\right) - w_i(T)\log(S_i(T))+w_i(t_0)\log(S_i(t_0)) + \int_{t_0}^{T}d\log(S_i(t))w_i(t) \right].
\end{align}

\noindent Setting $t_0=t$,

\begin{align} \nonumber
V(T) = V(t)\prod_{i=1}^{n}\left(\frac{w_i(T)}{S_i(T)}\right)^{w_i(T)}\left(\frac{S_i(t)}{w_i(t)}\right)^{w_i(t)}e^{\int_{t}^{T}w_i(t)d\log(S_i(t))}.
\end{align}

\noindent Using the payoff function in \eqref{eq:gen_payoff},

\begin{align} \nonumber
G(T)=V(t)\prod_{i=1}^{n}\left(\frac{S_i(t)}{w_i(t)}\right)^{w_i(t)}e^{\int_{t}^{T}w_i(t)d\log(S_i(t))}.
\end{align}

\noindent Noting that \eqref{eq:gen_payoff} also implies

\begin{align} \nonumber
V(t) &= G(t)\prod_{i=1}^{n}\left(\frac{w_i(t)}{S_i(t)}\right)^{w_i(t)}
\end{align}

\noindent gives

\begin{align} \nonumber
G(T)=G(t)\prod_{i=1}^{n}e^{\int_{t}^{T}w_i(t)d\log(S_i(t))},
\end{align}

\noindent as desired.

\subsection{Proof of Proposition \ref{prop:time_weights}}

\noindent Expanding in \eqref{eq:gen_dynamic_payoff}, we have

\begin{align} \nonumber
G(T) &=G(t)\prod_{i=1}^{n}e^{(r-\frac{\sigma_i^2}{2})\int_{t}^{T}w_i(t)dt + \sigma_i\int_{t}^{T}w_i(t)dW-i(t)} \\ \nonumber
&=G(t)e^{r(T-t) - \sum_{i=1}^{n} \frac{\sigma_i^2}{2}\int_{t}^{T}w_i(t)dt + \sigma_i\int_{t}^{T}w_i(t)dW_i(t)}.
\end{align}

\noindent   Taking expectations, we obtain

\begin{align} \nonumber
\tilde{\mathbb{E}}\left[e^{-r(T-t)}G(T)| \mathcal{F}(t) \right] &= \tilde{\mathbb{E}}\left[G(t)e^{-r(T-t)}e^{r - \sum_{i=1}^{n} \frac{\sigma_i^2}{2}\int_{t}^{T}w_i(t)dt + \sigma_i\int_{t}^{T}w_i(t)dW_i(t)} \right] \\ &=
G(t)e^{- \sum_{i=1}^{n} \frac{\sigma_i^2}{2}\int_{t}^{T}w_i(t)dt}\tilde{\mathbb{E}}\left[e^{\sigma_i\int_{t}^{T}w_i(t)dW_i(t)}| \mathcal{F}(t)  \right]. \label{eq:cl3}
\end{align}

\noindent Following the process outlined in \S\ref{sec:lemma_bms} now define the processes 

\begin{align} \nonumber
\sigma_P(t)=\sqrt{\sum_{i=1}^{n}w_i^2(t)\sigma_i^2 + \sum_{i \neq j} w_i(t) w_j(t)\sigma_i\sigma_j\rho_{ij}}
\end{align}

\noindent and

\begin{equation} \nonumber
Z_P(t)=\sum_{i=1}^{n}\int_{0}^{t}\frac{w_i(t)\sigma_{i}}{\sigma_P(t)}dW_j(u).
\end{equation}

\noindent where $Z_P(t)$ is a Brownian motion. Equation \eqref{eq:cl3} can now be written as

\begin{align} \nonumber
\tilde{\mathbb{E}}\left[e^{-r(T-t)}G(T)| \mathcal{F}(t) \right] &=
G(t)e^{- \sum_{i=1}^{n} \frac{\sigma_i^2}{2}\int_{t}^{T}w_i(t)dt+\int_{t}^{T}\frac{\sigma_P^2(t)}{2}dt} \\ &=
G(t)e^{\sum_{i=1}^{n}\frac{\sigma_i^2}{2}\int_{t}^{T}[w_i^2(t)-w_i(t)]dt + \frac{1}{2}\sum_{i \neq j}\sigma_i \sigma_j\rho_{ij} \int_{t}^{T}w_i(t)w_j(t)dt }, \nonumber
\end{align}

\noindent as desired.

\subsection{Proof of Proposition \ref{prop:target_weight}}

\noindent We seek a solution for $w(x,t)$ that satisfies

\begin{equation} \nonumber
G(0)e^{\int_{0}^{t}w^{*}(x,s)d\log(x)}=g(x,t)
\end{equation}

\noindent with initial condition $g(S_\alpha(0))=G(0)$. This is equivalent to

\begin{align} \nonumber
\int_{0}^{t}w^{*}(x,s)d\log(x)=\log{\frac{g(x,t)}{G(0)}},
\end{align}
which is solved by
$$
w^{*}(x,t) = \frac{d \log(g(x,t))}{d \log(x)}=  \frac{xg_x(x,t)}{g(x,t)}.
$$

\subsection{Proof of Corollary \ref{prop:deriv_target}}

The proof is identical to that of Proposition \ref{prop:target_weight}, except that we replace $x$ by $z(x)$ and $w(x,t)$ by $w(z(x))$.

\bibliography{references}
\bibliographystyle{acm} 

\end{document}